\documentclass[final,3p,times,twocolumn]{elsarticle}
\usepackage{graphicx}
\usepackage{amssymb}
\journal{Physics Letters B}

\newcommand{\be}{\begin{equation}}
\newcommand{\ee}[1]{\label{#1} \end{equation}}
\newcommand{\ba}{\begin{eqnarray}}
\newcommand{\ea}[1]{\label{#1} \end{eqnarray}}
\newcommand{\nl}{\nonumber \\}

\begin{document}

\begin{frontmatter}


 \ead{karoly.uermoessy@cern.ch}

\title{Cooper-Frye Formula and Non-extensive Coalescence at RHIC Energy}

 \author{K.~\"Urm\"ossy$^{a,b}$ and T.~S.~Bir\'o$^b$}
 \address[label1]{Roland E\"otv\"os University, Department of Theoretical Physics\\
P\'azm\'any P\'eter s\'et\'any 1/A, 1117 Budapest, Hungary}
 \address[label2]{KFKI Research Institute for Particle and Nuclear Physics of the Hungarian Academy of Sciences\\ Konkoly Thege Miklós \'ut 29-33, H-1121 Budapest, Hungary}

%%\author{K.~\"Urm\"ossy\corref{cor1} and T.~S.~Bir\'o\corref{cor2}}

%%\address{}

%%\cortext[cor1]{MTA KFKI RMKI, ELTE}
%%\cortext[cor2]{MTA KFKI RMKI}

\begin{abstract}
Transverse spectra are calculated for various types of hadrons stemming from Au Au collisions at $\sqrt{s}=200$ GeV. We utilize a quark recombination model based on generalized Boltzmann-Gibbs thermodynamics for local hadron production at various break-up scenarios.

%Using the Cooper-Frye formula transverse spectra are calculated for various types of hadrons coming from a locally thermal non-extensive quark-gluon plasma. The $\sqrt{s}=$200 GeV AuAu RHIC transverse spectra can be described well by the Boltzmann distribution under 2 GeV. This limit can be extended up to 5 GeV using the non-extensive Tsallis distribution.
\end{abstract}

\begin{keyword}
%%keywords here, in the form: keyword \sep keyword
non-extensive \sep coalescence \sep recombination \sep Cooper-Frye


\end{keyword}

\end{frontmatter}



\date{\today}

\section{Introduction}
\label{sec:int}
Hydrodynamics turned out to be a successful tool in describing the spacetime evolution of the hot dense quark-gluon plasma (QGP) created in heavy-ion collisions (HIC) until rehadronisation \cite{bib:cooperfrye}-\cite{bib:Heinz}.
Various types of quark recombination models based on Boltzmann-Gibbs (BG) thermodynamics \cite{bib:Levai}-\cite{bib:DukeUni5} have been used with success to reproduce the soft part ($p_T<2$ GeV) of transverse hadron spectra observed in Au Au collisions at $\sqrt{s}=200$ GeV \cite{bib:phenix1}-\cite{bib:star4}.
However in a QGP at hadronization there are long ranged interactions. Due to these, non-additive effects may not become negligible in the thermodynamical limit. The ratio of non-additive to additive terms in general scales like $surface \times force\, range / volume$. This quantity may not vanish in the limit of large volumes. In this case the BG thermodynamics has to be generalized \cite{bib:next13}.\\
Recent RHIC, Tevatron \cite{bib:teva} and LHC \cite{bib:cms} transverse hadron spectra are well fitted by the distribution, $dN/p_T\,dp_T\,dy=A(1+p_T/p_0)^{-n}$, at $p_T<5$ GeV. This formula has also been successfully applied in statistical and financial models.
There are various attempts to clarify the underlying physics that may result in such distributions. In works \cite{bib:next1}-\cite{bib:noisy} among others superstatistics, non-additive entropy or energy formulas due to long range correlations or interactions, fractal filling patterns in the phase space are discussed. However the microdynamical emergence of such a hadron distribution in a HIC is still not yet presented in a convincing calculation.
In the stringy \cite{bib:stringy}, and in the distributed mass  \cite{bib:distmassqm} quark coalescence model, quarks are assumed to be thermally distributed with the energy distribution
\be
 f(E)=A[1+(q-1)E/T]^{-1/(q-1)}.
\ee{hTsall}
Here T is the temperature of the QGP, q is a parameter showing the deformation of the BG thermodynamics. Sudden coalescence of n quarks results in hadron distributions of the cut power-law form with scaled parameters
\be
 T_H=T_q,\qquad n(q_H-1)=(q_q-1).
\ee{CoalRule}
It is also an open question whether the created matter reaches thermalization in the deconfined phase or it thermalizes in the hadronic phase, if at all.\\
In this paper we show that various hadron distributions all point out the hadron energy in the comoving frame as scaling variable. We assume that hadrons are produced at an instant propertime and the image of this distribution in the detector can be calculated via the Cooper-Frye (CF) formula. This calculation fits well to all types of transverse hadron spectra measured in $\sqrt{s} =$ 200 GeV AuAu collisions up to $p_T=5$ GeV.

\section{Transverse Spectrum Via the Cooper-Frye Formula}
\label{sec:hyp}
From the local equilibrium hadron distribution the transverse spectrum is calculated via the CF formula
\be
 E\frac{d^3N}{dp^3} = \frac{dN}{p_Tdp_Tdy d\varphi} =
 \int F(u^i p_i) \: p_i d^3\sigma^i.
\ee{hyp1}
with $p^i$ four-momentum of the detected particle, $u_i(x)$ local flow four-velocity (normalized to $u_iu^i=+1$) and $x^i$ spacetime coordinates, parametrized as
\ba
 x^i &=&  (\tau\,{\rm ch}\zeta, \: \tau\,{\rm sh}\zeta, \: r\cos\alpha, \: r\sin\alpha), \nl 
 u^i &=&  (\gamma\,{\rm ch}\eta, \: \gamma\,{\rm sh}\eta, \: \gamma v\cos\phi, \: \gamma v\sin\phi) ,\nl
 p^i &=&  (m_T\,{\rm ch}y,\:  m_T\,{\rm sh} y,\:  p_T\cos\varphi,\:  p_T\sin\varphi) .
\ea{hyp2}
Here $\tau=\sqrt{t^2-z^2}$ is the longitudinal proper time, $r$
the transverse (cylindrical) radius, $\zeta=\frac{1}{2}\ln\frac{t+z}{t-z}$
the longitudinal coordinate rapidity (hyperbolic arc angle) and $\alpha$ the
asimuthal angle belonging to the spacetime point $x^i$.
Similarily $v$ is the transverse flow velocity, $\gamma=1/\sqrt{1-v^2}$ the associated
Lorentz factor, $\eta$ the longitudinal rapidity of the flow and $\phi$ is its asimuthal angle. Finally $p_T$ is the detected transverse momentum,
$m_T=\sqrt{m^2+p_T^2}$ the corresponding transverse mass, while $y$ the observed longitudinal rapidity, $E=m_T\cosh(y)$ the energy and $\varphi$ the asimuthal angle of the detected paticle. This way the integration measure in (\ref{hyp1}) is given by
\ba
 p_id^3\sigma^i &=& m_T\cosh(y-\zeta) \, \tau r \, d\zeta dr d\alpha \nl
 &+&m_T\sinh(y-\zeta) \,  r \, d\tau  dr  d\alpha \nl
 &+& p_T\cos(\varphi-\alpha) \,  \tau r \, d\tau  d\zeta  d\alpha \nl
 &+&p_T\sin(\varphi-\alpha) \,  \tau \, d\tau  d\zeta  dr .
\ea{hyp3}
Considering a flow, constant and radial in the transverse plane ($\phi=\alpha$, v=const.) and scaling in the beam direction ($\eta=\zeta$), the energy in the comoving frame is
\ba
E^{co}&=&u^i p_i=\nl
&=&\gamma(m_T \cosh(y-\zeta) - v p_T \cos(\varphi-\alpha)) .
\ea{hyp4}
Assuming an instant hadronization at a constant longitudinal proper-time, $\tau_b$, and a continuous evaporation of hadrons at the cylindrical surface, the measured spectrum splits to four components:
\be
 E\frac{d^3N}{dp^3} = Y_0 + Y_1 + Y_2 + Y_4 .
\ee{hyp5}
These are
\ba
 Y_0 &=& \tau_b \int_0^{2\pi}\limits d\alpha \, \int_0^R\limits dr r \int_{-z}^{+z}\limits d\zeta \: m_T \: \times \nl && \qquad\times\:  \cosh(y-\zeta) F(\tau_b,\zeta,r,\alpha) ,\nl
 Y_1 &=& \int_0^{2\pi}\limits d\alpha \, \int_0^R\limits dr r \int_{0}^{\tau_b}\limits d\tau \,
 	m_T\:\times \nl && \qquad \times\: \bigl( \sinh(y-z)F(\tau,+z,r,\alpha) \nl
             && \qquad\qquad - \sinh(y+z)F(\tau,-z,r,\alpha)  \bigr) ,\nl
 Y_2 &=& R \int_0^{2\pi}\limits d\alpha \,  \int_{-z}^{+z}\limits d\zeta \, \int_{0}^{\tau_b}\limits d\tau \, \tau \,
	p_T \: \times \nl && \qquad\times\:  \cos(\varphi-\alpha) F(\tau,\zeta,R,\alpha )  ,\nl
 Y_3 &=&  \int_0^R\limits dr \int_{-z}^{+z}\limits d\zeta \,\int_{0}^{\tau_b}\limits d\tau \,\tau \: p_T \:\times \nl
       &&\qquad\times\: \bigl( \sin(\varphi-2\pi)F(\tau,\zeta,r,2\pi) \nl
       &&\qquad\qquad - \sin(\varphi)F(\tau,\zeta,r,0) \bigr) .\nl
\ea{hyp6}
Here $Y3=0$ for $F(E^{co})$ is periodic in $\alpha$, and $Y1\mapsto0$ taking the $z\mapsto \infty$ limit. This way the spectrum can be written in a more convenient form, i.e. as a sum of a volume and a surface term. After a shift in the integration variables $\zeta\rightarrow\zeta+y$ and $\alpha\rightarrow\alpha+\varphi$ the transverse spectrum takes the form
\ba
E\frac{d^3N}{dp^3}&=&A\int_{-\infty}^{\infty}\limits d\zeta \int_0^{2\pi}\limits d\alpha\:F(-\beta E^{co})\:\times\nl
     &&\times\:   (\xi\:m_T\cosh\zeta + (1-\xi)\:p_T\cos\alpha)
\ea{hyp7}
with
\ba
A\,\xi&=&2\pi\int_0^R\limits \, r \tau(r) dr = \pi\, R^2 \tau_b,\nl
A\,(1-\xi)&=&2\pi\int_{\tau(0)}^{\tau(R)}\limits \, r(\tau) \tau d\tau = \pi\, \tau^2_b\, R.
\ea{hyp8}
In the first integral of (\ref{hyp8}) $\tau(r)=\tau_b$, in the second $r(\tau)=R$. Assuming the BG hadron distribution $F(-\beta E^{co})=\exp(-\beta E^{co})/Z(\beta)$ we obtain
\ba
E\frac{d^3N}{dp^3}&=&A\bigl(\xi\: m_T K_1(\beta\gamma m_T) I_0(\beta \gamma v p_T)\nl
   &+&(1-\xi)\: p_T K_0(\beta\gamma m_T) I_1(\beta \gamma v p_T) \bigr).
\ea{hyp9}
This result is a decreasing nearly exponential function of $E^*=\gamma(m_T-v\,p_T)$ above $p_T=$ 2 GeV. However, using the hadron distribution (\ref{hTsall}) we utilize its Euler-Gamma representation \cite{bib:beck1}-\cite{bib:beck6}. Here $c=1/(q-1)$ and
\ba
\frac{dN}{p_Tdp_Tdy}_{y=0}&=&\frac{1}{\Gamma(c)}\int _0^\infty\limits dw \;w^{c-1}
        \int_{-\infty}^{\infty}\limits d\zeta \int_0^{2\pi}\limits d\alpha\:\times\nl
        &\times&(\xi\:m_T\cosh\zeta + (1-\xi)\:p_T\cos\alpha)\bigskip  \nl
        &\times&\exp \{ -w[1+(q-1)\beta E^{co}]\}.
\ea{intspe}
The integration over $\zeta$ and $\alpha$ leads to the spectrum
\ba
\frac{dN_{y=0}}{p_Tdp_Tdy}&=&A\frac{\xi\,m_T\,G_0(p_T)+(1-\xi)p_T\,G_2(p_T)}{(1+(q-1)\beta E^*)^{\,c}}.\nl &&
\ea{spf}
It differs from the results presented in \cite{bib:stringy} by the terms $m_T G_0$ and $p_T G_2$. The factors $G_i$ show, in fact, a very mild $p_T$ dependence.
\ba
G_{0,2}(p_T)&=&
\sum _j \frac{w_j z^{c-1}_j}{\Gamma(c)}e^{a_j-b_j}K_{1,0}(a_j)I_{0,1}(b_j)\:,\nl
a_j&=&\frac{(q-1)\beta\gamma m_T}{1+(q-1)\beta E^*}z_j\:, \nl
b_j&=&\frac{(q-1)\beta\gamma vp_T}{1+(q-1)\beta E^*}z_j\:
\ea{G}
with $w_j=z_j/(n-1)^2L(n+1,z_j)^2$, $L(n,x)$ being the Laguerre polynomial, and $L(n,z_j)=0$.
\section{Results}%------------------------------------------------------------------------
\begin{figure}[hp]
 \centering
 \includegraphics[width=0.92\textwidth]{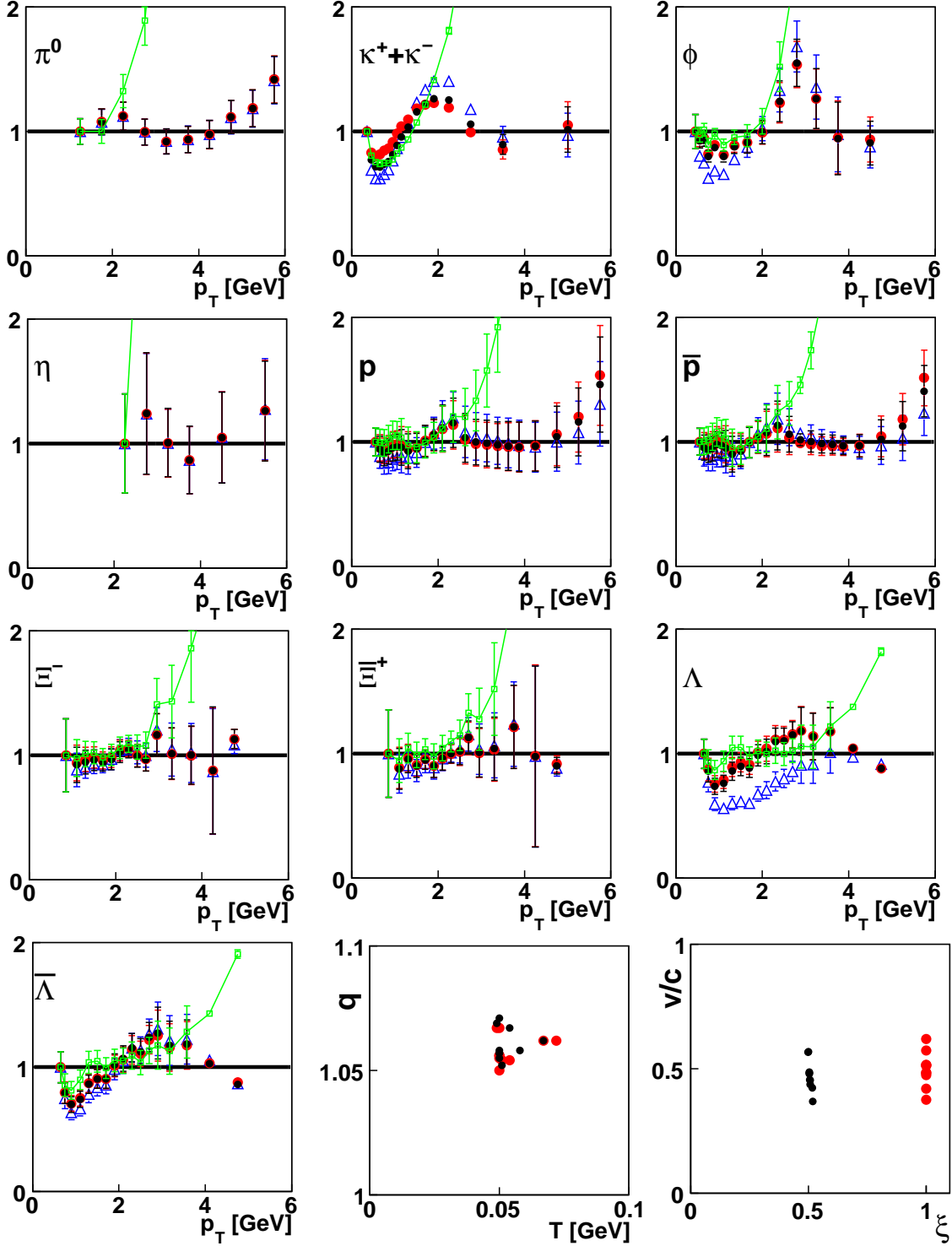}
 \caption{panels 1-10: Ratios of RHIC 200 GeV AuAu data and calculated spectra. Boxes: results of BG thermodynamics; triangles, dark and light spots: results of the hadron distribution (\ref{hTsall}) with $\xi=0,\,0.5,\,1$ respectively; Panels 11-12: fit parameters.}
 \label{fig:ratios}
\end{figure}

\label{sec:rhic}
The first 10 panels in Figure \ref{fig:ratios} show the ratios of meassured and calculated transverse hadron spectra $dN/p_T\,dp_T\,dy$ at mid rapidity ($y=0$).
Boxes denote results stemming from BG thermodynamics, triangles, dark and light dots denote the results obtained from the generalized hadron distribution (\ref{hTsall}). Triangles belong to $\xi=0$, dark dots to $\xi=0.5$ and light dots to $\xi=1$. This variable reflects the surface to volume term ratio in the hadronization formula (\ref{hyp1}). It is clear that the applicability of the BG distribution is limited to a very narrow range, up to 2 GeV, while the distribution (\ref{hTsall}) fits better than a factor 1.5 up to 5-6 GeV for all hadron types. Furthermore it is likely that most of the particles are produced in the instant hadronization process at $\tau_b$. The data/theory ratios are closest to one for $\xi=1$ which is taken in most calculations in the literature.
Panel 11 on Figure \ref{fig:ratios} shows the extracted values for the temperature, $T$, against those of the non-extensivity parameters ($q_H$). In panel 12 on Figure \ref{fig:ratios} transverse flow velocities ($v/c$) are plotted vs. the ratios of surface and volume yields ($\xi$). The cases $\xi=0.5$ and $\xi=1$ are depicted by dark and light dots respectively. The fitted temperatures and non-extensivity parameters are close for all hadrons ($T=51\pm10$ MeV, $q=1.062 \pm 7.65\times10^{-3}$). We conclude that the matter created in these HICs is thermalized according to energy shells, although the energy distribution is not the BG exponential \cite{bib:last1,bib:last2}.



\end{document}